\documentclass[letterpaper]{article} % DO NOT CHANGE THIS
\usepackage{aaai24}  % DO NOT CHANGE THIS
\usepackage{times}  % DO NOT CHANGE THIS
\usepackage{helvet}  % DO NOT CHANGE THIS
\usepackage{courier}  % DO NOT CHANGE THIS
\usepackage[hyphens]{url}  % DO NOT CHANGE THIS
\usepackage{graphicx} % DO NOT CHANGE THIS
\usepackage{natbib}  % DO NOT CHANGE THIS AND DO NOT ADD ANY OPTIONS TO IT
\usepackage{caption} % DO NOT CHANGE THIS AND DO NOT ADD ANY OPTIONS TO IT
\usepackage{algorithm}
\usepackage{algorithmic}

\usepackage{newfloat}
\usepackage{listings}
\DeclareCaptionStyle{ruled}{labelfont=normalfont,labelsep=colon,strut=off} % DO NOT CHANGE THIS
\floatstyle{ruled}
\newfloat{listing}{tb}{lst}{}
\floatname{listing}{Listing}

\title{Annotator in the Loop: A Case Study of In-Depth Rater Engagement to Create a Bridging Benchmark Dataset}
\author{
Sonja Schmer-Galunder\textsuperscript{\rm 1},
Ruta Wheelock\textsuperscript{\rm 2},
Scott Friedman\textsuperscript{\rm 2},
Alyssa Chvasta\textsuperscript{\rm 3},
Zaria Jalan\textsuperscript{\rm 3} and
Emily Saltz\textsuperscript{\rm 3} \\
}
\affiliations{
\textsuperscript{\rm 1} University of Florida, Gainesville, FL, USA \\
\textsuperscript{\rm 2} SIFT, Minneapolis, MN, USA \\
\textsuperscript{\rm 3} Google Jigsaw, Mountain View, CA, USA \\
s.schmergalunder@ufl.edu, rwheelock@sift.net, sfriedman@sift.net, \\
achvasta@google.com, zaria@google.com, emilysaltz@google.com
}

\begin{document}

\maketitle

\begin{abstract}
With the growing prevalence of large language models, it is increasingly common to annotate datasets for machine learning using pools of crowd raters. However, these raters often work in isolation as individual crowdworkers. In this work, we regard annotation not merely as inexpensive, scalable labor, but rather as a nuanced interpretative effort to discern the meaning of what is being said in a text. We describe a novel, collaborative, and iterative annotator-in-the-loop methodology for annotation, resulting in a 'Bridging Benchmark Dataset' of comments relevant to bridging divides, annotated from 11,973 textual posts in the Civil Comments dataset. The methodology differs from popular anonymous crowd-rating annotation processes due to its use of an in-depth, iterative engagement with seven US-based raters to (1) collaboratively refine the definitions of the to-be-annotated concepts and then (2) iteratively annotate complex social concepts, with check-in meetings and discussions. This approach addresses some shortcomings of current anonymous crowd-based annotation work, and we present empirical evidence of the performance of our annotation process in the form of inter-rater reliability. Our findings indicate that collaborative engagement with annotators can enhance annotation methods, as opposed to relying solely on isolated work conducted remotely. We provide an overview of the input texts, attributes, and annotation process, along with the empirical results and the resulting benchmark dataset, categorized according to the following attributes: Alienation, Compassion, Reasoning, Curiosity, Moral Outrage, and Respect.
\end{abstract}

\section{Introduction}

Current annotation practices pose broad issues that diminish data quality, reduce the diversity of human perspectives \cite{carrion2022few, miceli2020between}, and harm human well-being \cite{diaz2022crowdworksheets}. Within the Machine Learning community, a focus on data-hungry models and optimization for inter-rater reliability (IRR) has led to a focus on data quantity over data quality \cite{paullada2021data}. Quality metrics that do exist may be insufficient: while commonly-used metrics like IRR capture concept agreement within models, such metrics do not provide insight into the validity of the measured concepts themselves.  

A high-integrity annotation methodology is especially critical when labeling sensitive, nuanced, or complex linguistic attributes, ranging from qualities associated with destructive conflict (e.g., moral outrage, alienation or dehumanization) to constructive conflict to bridge divides (e.g., reasoning, curiosity, respect), as detailed in Table 1. Recent datasets of [un]healthy discourse have focused on toxicity, e.g., personal attacks \cite{zhang2018conversations}, bridging attributes and moral foundations \cite{kolhatkar2020classifying}, e.g., constructiveness \cite{saltz2024re}, or positive emotions, e.g., amusement \cite{paletz2023social}. Past efforts to do this have relied on ratings from research assistants or on crowd-rater platforms like CrowdFlower or Figure Eight, as is standard in the industry, where isolated annotators may label attributes based on their own intuitive perspectives, without room for horizontal discussions with others, agency to influence guidelines, and without occasional check-ins \cite{miceli2022data}. Annotators are not given opportunities to form intersubjective agreement, receive feedback, or may forget to exercise desired strategies, e.g., considering the perspective of the author of a post, even if they disagree with them \cite{miceli2022data, miceli2020between}. This can be particularly problematic for polarizing political topics.

Rater annotations of these attributes often draw on social and cultural knowledge of the context they are embedded in, including individual interpretations of embedded values, norms, or virtues. Attributes may have well-founded, theory-driven, academic definitions that researchers share as guidelines for annotators—and these may seem agreeable to annotators—but annotators also have intuitive ideas about these attributes, drawn from their life experiences, that they employ while annotating. Previous work has found that annotators have no way of providing feedback or asking questions, which has led to frustration among them \cite{diaz2022crowdworksheets, berg2015income}.

Our annotation process, described below, uses a novel approach that allows annotators to debate and deliberate collaboratively, refine the attribute definitions, discuss examples, and incorporate their intuitive ideas into the collective understanding of the concepts. This type of annotation stands in contrast to current practice, where annotators work in isolation. While we may strive for high agreement and “objectivity" in an annotation study, we question whether this is best achieved through ratings by isolated individuals. According to Haslinger’s theory of individual “situatedness,” multiple, differently situated observers of a phenomenon are desirable in order to increase objectivity \cite{haslanger2012resisting}. In other words, the collective understanding among annotators with diverse viewpoints influences the deep, human work of labeling these attributes.

The paper describes our novel annotator-in-the-loop annotation process: (1) we selected six attributes to annotate: Alienation, Compassion, Reasoning, Curiosity, Moral Outrage, and Respect (see Table 1); (2) we grounded the attributes in social science theory as well as pragmatic annotator ideas; (3) we engaged in iterative annotation and discussion sessions; and (4) we evaluated and trained classifiers on the resulting \textit{Prosocial Conversations for Bridging Benchmark Dataset} available on github.\footnote{
\textit{Prosocial Conversations for Bridging Benchmark Dataset}: https://github.com/conversationai/prosocial-comments-bridging-benchmark/}
Our quantitative results provide evidence that our annotator-in-the-loop methodology results in higher inter-rater reliability than traditional annotation methods, and our qualitative results suggest that collective annotator engagement is more desirable than traditional approaches.

\subsection{Shortcomings of Annotation Practices}

One drawback of current annotation practices is that they may not rigorously validate diverse conceptual understandings and interpretations of raters. Concepts are generally human-constructed social categorizations, where annotators make judgments and inferences about which data represents certain concepts \cite{mccrae2005universal}.

For example, a dataset containing racism or other sensitive topics annotated only by people from a targeted cultural background—or only from a non-targeted background—could yield an identity gap in understanding the cultural nuance and context, leading to poorly curated, biased datasets. On the other hand, researchers and practitioners have explored frameworks for better documenting crowdworker labeling processes to include factors like labor conditions and cultural background that are not traditionally released along with datasets \cite{diaz2022crowdworksheets}. Others have sought ways to quantify "annotator identity sensitivities" that may affect rating patterns \cite{10.1145/3531146.3533216}. While such work helps to document potential issues, it is also crucial to understand practical ways to address them in real-world applications.

\subsection{Toward Collective Conceptual Understanding}

Many recent AI models are built on large "benchmark" datasets such as ConceptNet \cite{speer2017conceptnet}, WordNet \cite{miller1995wordnet}, ATOMIC \cite{sap2019atomic}, ImageNet \cite{deng2014scalable}, and others. For instance, ImageNet associated 14 million images from the Internet with WordNet's twenty thousand categories, using Amazon Mechanical Turk \cite{crawford2021atlas}. Annotations were carried out cheaply and without human oversight, yielding general, static categorical judgments derived from hierarchical taxonomies of concepts. This focus on data quantity over quality has led to both data bias and model bias \cite{roselli2019managing}.

To address those shortcomings we have used social theories \cite{civilcomments, kolhatkar2020classifying} to inform concept definitions and we have adopted a more deliberative approach to the data annotation process, allowing annotators to engage in discussions around their collective concept understanding. Thus, a ``correct" label may not be one that strictly aligns with static instructions for a predefined concept, but rather one that results from an interpretive, deliberative process. This means that the outcome of the labeling process should more closely reflect the collective knowledge generated by the group rather than the isolated judgments of individual annotators. It should be noted that many annotators have already come up with informal ways of discussing their various interpretations within their communities, although without official oversight or the ability for those discussions to influence rater guidelines (e.g. online forums for Mechanical Turk crowdworkers \footnote{https://www.mturkcrowd.com/}$^,$\footnote{https://www.reddit.com/r/mturk/}). While it may be useful to enable such discussions for annotators, we have taken a more proactive approach, and included human oversight from the beginning. 

\subsection{Rationale for this Work}

Aroyo et al. \cite{aroyo2015truth} have proposed that we may need to move away from the antiquated ideal of an objective truth and instead embrace the idea of “crowd truth” or a \textit{“new theory of truth that rejects the fallacy of a single truth for semantic interpretation, based on the intuition that human interpretation is subjective and that measuring annotations on the same objects of interpretation across a crowd will provide a useful representation of their subjectivity and the range of reasonable interpretations.”}

The objective of our annotation study was to enhance data quality by creating a feedback loop with annotators, allowing for the joint refinement of attribute interpretation. We acknowledged the subjectivity inherent in understanding social concepts and sought to have annotators adopt the perspective of the post or document's author, considering their emotions, feelings, and intentions rather than imposing their own. Recognizing that absolute objectivity is neither achievable nor desirable, we accepted that the annotators might not always perfectly interpret the author's thoughts and emotions. Empathic accuracy, which refers to the ability to correctly infer others' emotions, varies among individuals, as does the perceived intensity of attributes within the text. Drawing on hermeneutics and "fusion of horizons", where understanding emerges through the interaction of different perspectives \cite{krajewski2003gadamer}, we encourage annotators to share their interpretations of attribute definitions and their strategies for evaluating these attributes. We sought to foster nuanced discussions and establish a new standard for prosocial discourse in annotation efforts.

\subsection{Human Oversight for Annotators}

High quality data in the domain of open online discourse means that annotations need to cover diverse perspectives, and consider the context and intentions of the author of a text (e.g. blog or comment).

Such high-integrity considerations can be easily communicated to annotators via training sessions. Annotators are able to exercise these considerations with \emph{human oversight} in the form of weekly feedback sessions including cognitive interviewing methods \cite{schwarz1996answering, campanelli1989role}.
The goal with our approach was to change the nature of the work from anonymous ``micro-workers'' operating in cyberspace isolation to one where workers can exchange their insight and experiences with a human mediator who routinely checks in with their peers. Collectively reviewing the rationale that supports a specific ``label'' for a nuanced example improves the working environment and data quality via group reflection and knowledge exchange.

In the following sections we describe the construction of the dataset, the selection of attributes, and the pool of annotators. We then describe the annotation process and present qualitative and quantitative results of our study, including inter-rater reliability and classification benchmarks. We conclude with a review of future work, limitations, and ethical considerations.

 \begin{table*}[ht!]
\footnotesize
\centering
\begin{tabular}{p{3.25in} p{2.75in}}
\textbf{Attribute \& Supporting Literature} & \textbf{Example Comments} \\ \hline

\textbf{Reasoning}: \emph{Makes specific or well-reasoned points to more fully understand the topic without disrespect or provocation.}

Previous work focuses on specific points and evidence \cite{kolhatkar2020classifying} and \emph{integrative complexity} to incorporate multiple perspectives into reasoning \cite{civilcomments}.
& 
\textbullet\ ``The bars on the windows of my neighbors that have lived through the transformation is proof positive. Not all urban revitalization is bad. Any city worth living in goes through urban revitalization, it is a positive sign of good city health.'' \\ \hline

\textbf{Curiosity}: \emph{Attempts to clarify or ask follow-up questions to better understand another person or topic.}

Supporting work on dialogue \cite{kolhatkar2020classifying}, \emph{behavioral complexity} involving explicitly asking questions designed to better understand a point of view \cite{civilcomments}, and collaboration \cite{de2021beg}.
& 
\textbullet\ ``Except the federal workers won’t be facing an income reduction, only a delayed payment. I wonder if that restricts them from accessing the program.''

\textbullet\ ``Is France sexist because Le Pen lost or did Le Pen lose because France is sexist? Or is it neither of those reasons? What if her gender had nothing to do with it...'' \\ \hline

\textbf{Respect}: \emph{Shows regard, deference, or courtesy towards views and beliefs different from their author's own.}

Literature on \emph{bridging systems} enabling mutual respect across divides \cite{ovadya2023bridging}, politeness \cite{danescu2013computational}, respect \cite{stroud2011niche}, and the interaction of [dis]respect with informativeness and persuasiveness \cite{napoles2017finding}.
& 
\textbullet\ ``Thanks, didn’t know there was a term for it. That’s how I handled the first and last episode I saw lol.''

\textbullet\ ``You're not the only one. I for one feel stupid for thinking 100t shouldn't be having problems with such an easy pool.'' \\ \hline

\textbf{Compassion}: \emph{Identifies with or shows concern, compassion, or support for the feelings/emotions of others.}

\emph{Supportiveness} is an emotional strategy \cite{wang2018s} and plays a role in prosocial language \cite{bao2021conversations}.
& 
\textbullet\ ``...but trust me when I say there are people here who'll do everything they can to look after you.''

\textbullet\ ``I am here to tell you that it really will work out okay! I have certainly felt like you before...'' \\ \hline

\textbf{Alienation}: \emph{Includes language that refers to an opposing individual or group as less human, inferior, and/or not fitting in within the norms of a social group or society.}

Describing groups or individuals as alien or less-than-human may persuade audiences to deny moral consideration they give to those who are in-grouped. Reducing this animosity is necessary for bridging \cite{ovadya2023bridging}.
& 
\textbullet\ ``Canadians virtue smug are signaling pathetic idiots''

\textbullet\ ``Zero class on all of them. While Tory is closing social housing units (talk about hitting the most vulnerable the hardest) ... while my hydro bill has a Liberal propaganda message on the outside (We have a plan for lowering electricity costs). Man, what a hypocrite.'' \\ \hline

\textbf{Moral Outrage}: \emph{Directs anger, disgust, or frustration at those who violate implicit/explicit ethical values or standards.}

Positive social feedback for outrage expressions encourages future outrage expressions \cite{brady2021social, brady2020mad}. Outrage-inducing content is more prevalent and potent online than offline \cite{salerno2013interactive}, despite end-users disliking its virality \cite{rathje2023people}. Over-perception of online moral outrage inflates beliefs about inter-group hostility \cite{brady2023overperception} and may stem from deceptive content \cite{carrasco2022fingerprints}.
& 
\textbullet\ ``Plagiarism sickens me because its motive is vanity.''

\textbullet\ ``I hope you get caught some day and have to pay for all of the damage. I don't know why you would possibly find this entertaining.'' \\ \hline

\end{tabular}
\caption{Target attributes, supporting motivation in the literature, and examples.}
\label{tab:attributes}
\end{table*}

\section{Dataset}

The dataset is composed of 11,973 comments from Civil Comments, which is a publicly available dataset of comments from independent and international news sites that were created from 2015-2017 \cite{borkan2019nuanced}. The data is labeled for six attributes: constructive, curiosity, respect, empathy, alienation, and moral outrage. Due to the low prevalence of these attributes and budget constraints, the data that was annotated was first scored by a proprietary model and then filtered by score to ensure a higher proportion of in-class comments. The data were sent to the annotators in three batches which allowed for iterative data sampling improvements as time went on, and later batches of the data constrain the minimum and maximum text length and limit the amount of text concerned with Canadian politics by dropping the comments containing the terms Trudeau and Canada. Additionally 698 examples were selected to include identity terms based on the Civil Comments identity labels defined in \citet{borkan2019nuanced}. 

\section{Methods}

In this section we describe the annotator selection criteria and process, how we decided on which prosocial concepts to annotate, and what type of qualitative and quantitative methods we implemented.

\subsection{Annotators}

We utilized a staffing agency to recruit candidates for annotation. We formed two groups of annotators, each consisting of four members. Annotators were hired part-time with a flexible schedule, working 17 hours per week for a duration of 10 weeks with a base salary of \$15 per hour. One of the groups consistently annotated above the minimum annotation speed with good agreement scores, so their hourly rate was increased to \$18 after a few weeks. Annotators received bonus payments after the project ended, proportional to their contributions to the project.

We collected anonymous demographic information at the end of the study and received responses from 7 annotators. See Appendix A for a summary of the demographic survey data from the annotators.

\subsection{Attribute selection}
\label{sec:selection}
To determine the six chosen concepts from a list of over 20 concepts the research team had put together, we conducted two pilot studies using the crowd-sourcing platform Prolific. Each study enlisted 40 crowd-source workers to evaluate the presence and intensity of attributes in a curated set of posts. The studies resulted in over 200 annotated posts each, with an average of five annotations per post. To measure agreement among annotators, we calculated Krippendorff's alpha. We used Krippendorff's alpha instead of Fleiss's Kappa to measure agreement because, in our pilot studies, we also rated the intensity of attributes, and kappa is not suited for ordinal data. The first pilot study yielded relatively low alpha scores, ranging from 0.09 to 0.25. However, after refining the definitions for the second pilot study, the alpha scores improved for almost all attributes, with the range from 0.12 to 0.43. The outcomes of these studies informed the choice of the concepts most useful in describing prosociality and enabled us to refine our definitions. Moreover, these studies provided data on the average speed per annotation, establishing a benchmark for future annotation efforts, and garnered feedback from crowd-workers on the study's design, attribute definitions, and data.

\subsection{Qualitative and Quantitative Methods}
Many crowd-sourced annotation projects rely solely on quantitative methods to evaluate annotation work, such as annotation speed, inter-rater reliability, correctness against a gold standard, or cost per annotated post. While these metrics are important, we propose the inclusion of qualitative methods. This study demonstrates how qualitative methods can be integrated at different stages of the annotation process, from interviewing candidates to training, weekly feedback meetings, and exit interviews.

\subsubsection{Qualitative Methods}

Any data annotation project begins with the selection of annotators. Instead of using anonymous crowd-sourced workers, often from the global South \cite{miceli2022data}, we hired individual US-based annotators from a temp worker agency using predefined selection criteria. We conducted 30-minute cognitive interviews \cite{castillo2013cognitive}, where we briefed them on the project's objectives. We sought candidates with relevant work experience, high proficiency in English, and the ability to articulate their thought process clearly when assigning attribute labels. During the interviews, we provided training examples and used a Think Aloud protocol, where annotators explained their reasoning for their choices.

Next, we incorporated qualitative methods into the training. Due to epistemological differences between the definition of an attribute (which we provided) and how individuals understood and interpreted these attributes in text examples, we included feedback from the annotators during training to update and improve concept definitions, making them more understandable. We considered this step essential to bridging the gap between theory and practical application. Training was conducted in groups of four annotators and one trainer. Each group underwent three training sessions, each lasting 1.5 hours, followed by 2.5 hours of individual annotation time after the final session. During the training sessions, we reviewed attribute definitions and examined examples. Annotators had the opportunity to annotate examples independently during each meeting, followed by discussions to address any agreements or disagreements on those examples. These discussions helped us understand what made our attribute definitions challenging to apply to the text and how different annotators approached these challenges. After the final training session, annotators were tasked with annotating as many examples as possible within 2.5 hours. This individual training work, not included in the dataset, aimed to familiarize the annotators with the annotation process and deepen their understanding of the attributes. Once training was complete, the definitions were not changed.

To evaluate progress and align annotator attribute interpretations as the annotation work continued, one of the researchers held weekly online meetings with each group of four annotators. During these meetings, we provided feedback in the form of quantitative results from their previous week's work and discussed examples they found challenging. As reported below, quantitative results helped us identify which attributes had the largest disagreements, but they did not explain why these disagreements occurred. Therefore, discussions among annotators and their collective sharing of their conceptual understandings were essential to understanding the reasons behind these disagreements. Initially, these meetings acted as an extension of the training. However, as annotators solidified their understanding of the attributes, the meetings shifted to serve primarily as mental health check-ins, maintaining motivation, and providing an opportunity for annotators to give feedback to the requester or ask questions.

After the annotation work was completed, we asked the annotators to fill out a project feedback survey and conducted individual 30-minute exit interviews with each annotator. The surveys and interviews helped us better understand how different parts of the project were perceived by the annotators, what challenges they experienced, and how we can improve the process in the future.

\subsubsection{Quantitative Methods}

We used multiple quantitative methods to measure agreement among annotators and track progress. Two main metrics were the reliability coefficient Krippendorff's alpha, used to measure annotator agreement, and annotation speed (posts per hour) to evaluate annotator efficacy.

To set expectations for annotation work, we established a minimum annotation speed (posts per hour) that annotators had to meet. Based on data from pilot studies, we set the minimum limit to 30 annotated posts per hour. The annotation speed is affected by many factors, including the length of posts, the number of attributes to annotate, the complexity, and the content of the text. One of the annotator groups consistently annotated more than 40 posts per hour with very good inter-rater reliability, so their hourly rate was increased, and the minimum annotation speed was set to 40 posts per hour after a few weeks of annotation work. Annotators exhibited a wide range of annotation speeds, with some consistently handling 60 posts per hour. To compensate for the disproportionate contributions to the project that were not reflected in the hourly rate, we issued bonus payments after the project ended based on each annotator's individual performance. Using speed as a measure of annotator performance is simple and straightforward; however, evaluating the quality of annotations is much more challenging.

\section{Results}

\begin{table*}[htb]
\footnotesize
\centering
\begin{tabular}{p{2.5in} p{3.5in}}
\textbf{Topic \& Summary} & \textbf{Qualitative Feedback (direct quotes)} \\ \hline

\textbf{Concept Definitions}: Annotators reported a duality between specificity of definitions and subjective interpretation into their own worldview.
& 
 ``Our group spent a lot of time debating on how we interpret the definitions which influenced how we rated the responses.''
 ``More specific attribute definitions had to be devised by annotators as the project went on.''
``Don't finalize the attribute categories and definitions till after the training week.''
 ``[...] many of them were either very limiting, or too broadly open to interpretation.'' \\ \hline

\textbf{Training Annotators}: Group training was paramount for task and concept understanding. This evolved over multiple meetings. Working with examples during training (including providing reference examples) was crucial.
& 
``[Instructions] were less clear in the beginning but as we progressed became more clearly defined as we walked through the examples with the project lead and the other annotators.''
 ``[It] took a few rounds before getting in the flow. Harder to work back to where our work and annotations were consistent with the team as a whole.'' \\ \hline

\textbf{Annotation as a Process}: Although annotators got better at their task, skills were still evolving and conceptual understanding was continuously updated several weeks after start.
& 
 ``Annotation should be a process not application (where each annotator decides how they interpret definitions).''
``The meetings were extremely helpful. Hearing my colleagues' approaches to annotating let me know that we were all, more or less, on the same page, and it helped me refine my own approaches to comments that were more difficult to label.'' \\ \hline

\textbf{Information Exchange between Annotators}: Weekly meetings—including cognitive interviewing and discussing examples with other annotators—were considered helpful in shaping the interpretative work. Information exchange and deliberate debates across viewpoints led to collective understanding.
& 
 ``Knowing how my colleagues were parsing through the comments and hearing their thinking about them helped me refine my own understanding of, and approach to, labeling the comments.''
``I very much needed to hear how they were thinking about their own approaches to annotating.''
 ``It helped when there was some disagreement on how to interpret the attribute definitions.'' \\ \hline

\textbf{Social Positions of Annotators}: Socio-economic background and culture either helped or hindered the interpretation of the textual content, especially when jargon was used, or annotators were not familiar with the politics of the country.
& 
``I did benefit from my Asian-American background and experience living abroad, which gave me a different perspective with texts that pertained to Hawaii or Japan.''
``Anything involving very far right politics was very difficult to remain unbiased.''
 ``Yes, not having much knowledge of Canadian politics or Catholic teachings hindered my understanding of some of the annotations.'' \\ \hline

\textbf{Mental Health of Annotators}: Toxic content causes distress in annotators and reduces the amount of work time annotators were willing to put in. Mixing negative content with positive content alleviated some of the mental drain.
& 
 ``Overall, I don't think it impacted my mental health in a major or long-lasting way. However, it is nice to have a set of data that had more compassionate texts or was more balanced across the various attributes.''
 ``Sometimes, the experience was very much disillusioning and demoralizing, and I would dread commencing my weekly [meeting].''
 ``Yeah, there was plenty of anti-queer content that bothered me, but honestly I'm used to seeing stuff like that online as part of participating in discourse that is personally important to me.'' \\ \hline

\end{tabular}
\caption{Annotation topics and representative examples of direct quotes.}
\label{tab:results}
\end{table*}

\subsection{Qualitative Results}

First, we observed a tension between predefined theoretical frameworks and the personal intentions of the authors. This tension arose from the dichotomy between specific versus general, broad concepts, and the debate over whether concepts should encompass multiple facets or if those facets should be treated as distinct concepts. The challenge of categorizing documents within a predefined framework often fails if annotators think of single facets within a category as defining features, or, on the contrary, if only a specific facet of a concept (presumably ill-defined) is the goal, but annotators may cast a wider net. For example, the category “compassion” may be interpreted as “perspective-taking” or “having empathy,” while empathy itself has a cognitive (what are the other person’s thoughts or beliefs) vs. emotional (what is the other person feeling) component. This conceptual ambiguity can lead to cognitive dissonance and frustration among annotators, as they struggle to align their nuanced understanding with rigid categorization schemes, echoing concerns raised in philosophy of language (e.g. \cite{wittgenstein1958blue}) and the challenges of defining complex concepts through strict taxonomies.

Second, we saw that initial and ongoing training is important. We observed that annotators continued to discover new challenges in interpreting attribute definitions weeks after the annotation work began. This indicates that some intricate nuances in the definitions require many hours of annotation work to surface. Current standard practice may keep this part at a minimum, but we found that ongoing training, including some flexibility with regards to re-defining attributes based on the data evidence, improves the understanding of the annotators \textit{and} yields better results (see Tables 1 and 2). Third, although most annotators improve with practice, we found that new data samples, or a change of data, yield new challenges in understanding (e.g., new socio-political context or new jargon of a specialized online community) and that consistency, including a human facilitator in the process, was considered immensely helpful and less isolating. This finding aligns with John Dewey's pragmatist philosophy \cite{miettinen2000concept}, which emphasizes experiential learning and the continuous adaptation of knowledge to new information, suggesting that such a dynamic training process can enhance annotators' engagement and psychological well-being by providing a more supportive and evolving learning environment.

Fourth, the opportunity to hear how others ascribed meaning to what was said by an author was an important exercise in perspective-taking and understanding different “ways of seeing” \cite{berger2008ways}. For example, an annotator from Europe may not have enough cultural knowledge about the local politics of Canada to understand the full meaning, subtleties or sarcasm. This point cannot be stressed enough, especially given that crowd-source workers often annotate data produced by the global north, while being situated in the global south \cite{miceli2022data}. It should be stressed that hermeneutics \cite{regan2012hans} emphasize the importance of historical and cultural context in interpretation, suggesting that without such contextual understanding, annotators may experience frustration or misinterpretation.

Finally, as has been previously observed in the literature, we noted the impact on the mental health of annotators encountering toxic text \cite{alemadi2024emotional, hao2023cleaning}. Having participants from marginalized or underrepresented groups annotate problematic data can affect the mental health of those individuals and needs to be addressed \cite{smart2024discipline}. In addition to such considerations, we also found that annotators benefited from deciding how many hours they were able to work on a task without feeling overwhelmed. Thus, ensuring that annotators are not subjected to alienating experiences underscores the importance of maintaining an ethical approach in managing their workload and mental health.

\begin{table*}[htb]
    \centering
    \begin{tabular}{r | c c c}
    & \multicolumn{3}{c}{\textbf{Inter-Rater Reliability}} \\ \hline
    \textbf{Attribute} & \textbf{Trad.} & \textbf{Novel} & \textbf{Diff}  \\ \hline
    constructive & 0.314 & 0.435 & 0.120 \\
    curiosity & 0.193 & 0.625 & 0.431 \\
    respect & n/a & 0.489 & n/a \\
    empathy & 0.217 & 0.442 & 0.225 \\
    alienation & 0.309 & 0.544 & 0.234 \\
    moral outrage & 0.300 & 0.532 & 0.232 \\ \hline
    \end{tabular}
    \caption{Inter-Rater Reliability scores comparing Traditional and Novel approaches to annotation.}
    \label{tab:irr_roc}
\end{table*}

\subsection{Quantitative Results}

For the quantitative results the data from this paper is compared against the baseline derived from previously labeled data using traditional crowd-sourcing labeling methods. This data is of similar size and was annotated by crowd raters based on a template, with no feedback component between the annotators and the annotation job creators. The label "Respect" was excluded in this analysis since we had no comparative data from previous work. Between this novel annotation method and the traditional labeling job, the Inter-Rater Reliability (IRR) was higher. Data annotated with the novel annotation method has an IRR that is on average 0.248 higher than traditionally annotated data (see Table 3.)T This is especially important because of the correlation between data with high IRR and model performance.

We calculated Krippendorff's alpha weekly for every attribute and analyzed how each annotator contributed to it. We observed that sometimes the data itself significantly influenced the agreement scores. For example, when a dataset for a rating period contained more toxic content, bridging attributes like ``reasoning'' or ``compassion'' were not present. This in turn lead to low agreement scores for those attributes as raters sought to identify any examples. Krippendorff's alpha indicates how well an annotator aligns with others but does not reflect the actual quality of the annotations.

\section{Summary and Conclusions}

In this paper we introduce a novel approach to data annotation based on building relationships with and between annotators, as well as with the requester of the work (e.g. facilitator or researcher), which functioned as a "human in the loop." We introduced qualitative methods that motivated knowledge exchange and reflective thinking on both an individual level, but also on a group level, thereby co-creating meaning and co-developing knowledge based on the exchange of rationales and individually situated thought processes (e.g. ``crowd truth''). We found that when looking at the work of annotation as a collective process of deliberate exchange of diverse ways of thinking, we can use qualitative methods to improve quantitative outcomes. This means that the proposed method can improve IRR, even when there is a lot of initial disagreement or diversity of viewpoints. This is a considerable step forward in annotation work, which traditionally focused on large quantities of data without giving data quality much consideration. With this work, we show that focusing on data quality (instead of quantity) using human facilitators, can yield better benchmark datasets, and subsequently better ML models. 

Further, when defining concepts and attributes for annotators, we found that social theory can inform the intended use of the dataset and the ML model. For example, developing a benchmark dataset for bridging, we built on our own previous work and conducted a review of the academic literature on prosocial discourse, including theories about politeness \cite{danescu2013computational,wu2013non}, moral justifications of harm \cite{friedman2021toward}, social regard towards others \cite{zheng2022towards,friedman2024debiasing}, empathy \cite{chen2021gender}, civility \cite{papacharissi2004democracy}, deliberate moderation \cite{friess2021collective}, and others listed under Table 1, allowing us to provide more tailored definitions. However, we also found that flexibility to theory-driven definitions is key and that those definitions need to be adjusted and refined to real-world examples.

Lastly, the research project addresses several ethical issues related to data annotation. We argue that a slow, deliberative process focused on quality over quantity ensures that even the initial step is undertaken responsibly. By understanding and listening to the individuals who perform the foundational work for ML models, we can improve the overall quality of the data and effectiveness of these models.

\section{Future Work and Limitations}

While we find that our iterative annotator-in-the-loop methodology addresses several ethical and societal issues, we acknowledge that this approach is both time-intensive and costly. It requires a considerable amount of additional work from a highly skilled professional to meet and work with annotators over several weeks. However, the cost may be worthwhile if the outcome is a more robust and accurate training dataset. Looking ahead, we propose moving away from a model of cheap and fast data labeling to a more deliberate, thoughtful, and slow process, mirroring aspects of “deep reading” from the anthropological and philosophical literature, at least for complex linguistic tasks. Such approaches may become even more important if we want to train generative models on human reasoning or other higher-order tasks.

An additional limitation is in the diversity of viewpoints. For our work, we only used U.S. persons. Future annotations would benefit from more diverse viewpoints (one annotator self-identified as left of center, none as center or to the right of it, see table below) and from other cultures. It should also be noted that in future work, technologists may want to ensure that annotators have the necessary socio-cultural and political context to better understand sarcasm and nuance.

Pieces of text are annotated out of its immediate context (what was said right before or after) as well as the broader political history and time. Therefore,  we propose the inclusion of relevant metadata, e.g. social context about the data being rated, or additional comments before and after the rated posts, to better inform interpretations and sense-making. And finally, to protect the mental well-being of annotators, we need to minimize their exposure to toxic content, or take appropriate precautions so as to avoid harm to annotators, such as the "HarmCheck" standard for handling and presenting harmful text in research \cite{kirk2022handling}.

\section*{Ethics Statement}
In this paper, we address several ethical issues related to the construction and annotation of datasets. While we cannot fully eliminate inherent biases within the datasets, we strived to minimize the introduction of biases during the annotation process. We ensured that annotators received fair compensation and had control over their work hours, addressing concerns about working conditions. To protect the mental well-being of annotators, we also minimized their exposure to toxic content.

Constructing a benchmark dataset for prosocial discourse is an ethical endeavor aimed at improving online conversations and positively impacting society. We believe our methodology and application represent significant improvements over current practices.

Nonetheless, ethical concerns persist regarding data collection, including the need for transparency, accountability, and privacy protection. Ensuring these involves removing any identifiable information to address these issues.

\section*{Acknowledgements}
We are deeply grateful to all of those who reviewed and enhanced this work with their thoughtful and rigorous feedback, including Eric Morley, Rachel Rosen, Tin Acosta, Daniel Borken, Drisana Iverson, Beth Gold and Jeffrey Sorenson.

\bibliography{aaai24}

 \vspace{5cm}

% No \appendix command in AAAI format
\section{Appendix}
% First Appendix Section
\section{Appendix A: Summary of Annotator Demographic Survey Data}
\label{sec:appendixA}

\subsection{Age Distribution}
\begin{itemize}
    \item 18-29 years: 2 respondents
    \item 30-39 years: 4 respondents
    \item 50-59 years: 1 respondent
\end{itemize}

\subsection{Gender Identity}
\begin{itemize}
    \item Man: 3 respondents
    \item Woman: 2 respondents
    \item Non-Binary: 2 respondents
\end{itemize}

\subsection{Sexual Orientation}
\begin{itemize}
    \item Heterosexual/Straight: 3 respondents
    \item Gay/Lesbian: 1 respondent
    \item Bisexual: 1 respondent
    \item Prefer not to say: 1 respondent
    \item Not Listed: 1 respondent
\end{itemize}

\subsection{Household Annual Income Before Taxes}
\begin{itemize}
    \item \$25,000 to \$50,000: 2 respondents
    \item \$50,000 to \$75,000: 2 respondents
    \item \$75,000 to \$100,000: 2 respondents
    \item Prefer not to say: 1 respondent
\end{itemize}

\subsection{Household Size}
\begin{itemize}
    \item All respondents live in single-person households.
\end{itemize}

\subsection{Ethnicity}
\begin{itemize}
    \item Caucasian: 3 respondents
    \item Asian: 1 respondent
    \item African-American: 1 respondent
    \item Not listed: 1 respondent
    \item Prefer not to say: 1 respondent
\end{itemize}

\subsection{Highest Attained Level of Education}
\begin{itemize}
    \item Bachelor's degree: 4 respondents
    \item Master's degree: 3 respondents
\end{itemize}

\subsection{Religion}
\begin{itemize}
    \item Catholicism/Christianity: 2 respondents
    \item Not applicable: 4 respondents
    \item Not listed: 1 respondent
\end{itemize}

\subsection{Political View}
\begin{itemize}
    \item Very Liberal: 3 respondents
    \item Slightly Liberal: 3 respondents
    \item Prefer not to say: 1 respondent
\end{itemize}

\subsection{Place of Birth}
\begin{itemize}
    \item All respondents were born in North America.
\end{itemize}

\subsection{City and State of Residence}
\begin{itemize}
    \item All respondents reside in the Minneapolis-Saint Paul metropolitan area, Minnesota.
\end{itemize}

% Second Appendix Section
\section{Appendix B: Annotator Feedback Survey Questions}
\label{sec:appendixB}

\begin{enumerate}
    \item How do you rate the task instructions?
    \item How do you rate the attribute definitions?
    \item How helpful was the training week in preparing you for the annotation work?
    \item How would you improve the training?
    \item How important was the interaction with other annotators in helping interpret the task?
    \item How often did your personal judgment and life experiences hinder you from being objective when evaluating the author's perspective? Please provide examples if that happened.
    \item Do you feel that your social background influenced your annotations in any way? If yes, explain.
    \item As someone who may have experienced online harassment, did this impact your approach to annotating potentially harmful content?
    \item Do you think that the content of the annotations was impacting your mental health?
    \item If applicable, do you think that being part of a certain (marginalized) group made you more sensitive to certain contents?
    \item Were you satisfied with the compensation provided for this annotation task?
    \item Were you satisfied with the working conditions provided for this annotation task?
    \item Did you encounter any content that was distressing or challenging to annotate? How did you handle this?
    \item Were you able to communicate effectively with the SIFT team if you had questions or needed clarifications?
    \item How helpful were the weekly meetings in understanding the task?
    \item How helpful were the weekly meetings in terms of letting others know how you feel?
    \item How do you perceive the relevance of this annotation task in relation to broader societal or community issues?
    \item Are there any ethical considerations or potential impacts of this dataset that you think should be addressed or monitored in its future use?
    \item Any other feedback or comments:
\end{enumerate}

\end{document}